\documentclass[journal,compsoc]{IEEEtran}

\usepackage{tikz}
\usetikzlibrary{arrows.meta, positioning, fit, calc}
\usetikzlibrary{automata, positioning}
\usepackage{comment}
\usepackage{algorithm}
\usepackage{algpseudocode}
\usepackage{amsmath,amssymb, mathtools}
\usepackage{subcaption}
\usepackage{xspace}
\DeclareMathOperator{\Lab}{Lab}     
\DeclareMathOperator{\Next}{Next}   
\DeclareMathOperator{\subsids}{subtree\_sids}
\DeclareMathOperator{\sid}{sid}

\usepackage{booktabs} 
\usepackage{multirow}
\usepackage[table]{xcolor}
\usepackage{balance}
\usetikzlibrary{arrows.meta, positioning}

\usepackage{tabularx}
\usepackage{tcolorbox}
\usepackage{arydshln}
\usepackage{tcolorbox}
\tcbset{colback=gray!5, colframe=black!70, boxrule=0.5pt, arc=2pt, left=4pt, right=4pt, top=4pt, bottom=4pt}
\usepackage{pifont}

\newcommand{\best}[1]{\multicolumn{1}{>{\columncolor{gray!40}}c}{#1}}
\newcommand{\second}[1]{\multicolumn{1}{>{\columncolor{gray!15}}c}{#1}}

\newcommand{\tool}{\textsc{EvoRec}\xspace}

\definecolor{prompttitle}{RGB}{60,60,60}
\definecolor{promptblue}{RGB}{30,60,180}
\definecolor{promptgray}{RGB}{245,245,245}
\definecolor{promptgreen}{RGB}{0,100,0}
\definecolor{promptpurple}{RGB}{110,0,180}

\makeatletter
\renewcommand{\paragraph}[1]{%
  \par\addvspace{1ex}%
  \noindent\hspace*{\parindent}%
  \textit{#1:}\ %
}

\newcommand{\add}[1]{\textcolor{black}{#1}}
\makeatother

\ifCLASSINFOpdf
\else
\fi
\begin{document}
%
\title{When Model Editing Meets Service Evolution: A Knowledge-Update Perspective for Service Recommendation}
%
%
%

%
%

\author{
Guodong~Fan,
Cuiyun~Gao*\thanks{*Corresponding author.},
Chun Yong Chong,
Lu~Zhang,
Jing~Li,
Jinglin~Zhang, 
and Shizhan~Chen
\IEEEcompsocitemizethanks{
\IEEEcompsocthanksitem
G. Fan is with the College of Information Science and Engineering, Shandong Agriculture and Engineering University, China.
C. Gao is with the College of Computer Science and Technology, Harbin Institute of Technology (Shenzhen), China.
C.Y. Chong is with School of Information Technology, Monash University Malaysia, Malaysia.
L. Zhang and S. Chen are with the College of Computer Science and Technology, Tianjin University, China.
J. Li is with the College of Computer Science and Technology, Shandong University of Technology, China.
J. Zhang is with the School of Control Science and Engineering, Shandong University, China.
}
}

\markboth{Journal of \LaTeX\ Class Files,~Vol.~14, No.~8, August~2015}%
{Shell \MakeLowercase{\textit{et al.}}: Bare Demo of IEEEtran.cls for IEEE Journals}
%



\maketitle

\begin{abstract}
The rapid evolution of software services poses substantial challenges to the design and implementation of effective recommendation systems. Traditional service recommendation approaches often rely on static representations and historical usage data, which are insufficient for adapting to the dynamic and evolving nature of service ecosystems. Recently, large language models (LLMs) have shown strong potential to overcome these limitations by leveraging rich contextual understanding. However, their practical use faces two major challenges: outdated service facts and invalid or redundant services. 
To address these issues, we propose \tool, an evolution-aware framework for service recommendation that leverages model editing in a locate–then–edit paradigm to incorporate updated service facts without costly retraining efficiently.
This allows the model to remain aligned with evolving service ecosystems. 
To address invalid service issues, we introduce a Finite Automata (FA)-based constrained decoding mechanism with deduplication, which enforces structural and semantic validity while eliminating repeated services. Experiments on real-world service datasets demonstrate that our framework consistently outperforms existing baselines, e.g., achieving an average relative improvement of 25.9\% in Recall@5. 
Moreover, under evolving service scenarios, our approach outperforms model fine-tuning approaches by 22.3\%, demonstrating strong adaptability to service evolution and providing a practical solution for service recommendation in dynamic ecosystems.

\end{abstract}


\IEEEpeerreviewmaketitle

\section{Introduction}
%
%
%
%

\IEEEPARstart{S}{ervice} computing has become the cornerstone of modern software ecosystems, enabling applications to be rapidly constructed by composing and reusing many heterogeneous services~\cite{siqueira2021service, fan2023service}. With the continuous expansion of service repositories, service recommendations have been extensively studied to assist developers in discovering, selecting, matching, and composing services that satisfy user requirements~\cite{fan2025user}. However, the service ecosystem is inherently dynamic and evolving, where services are frequently updated, deprecated, or replaced, while new services are constantly introduced~\cite{liu2021data}. This continuous service evolution poses significant challenges to existing recommendation approaches.

Traditional recommendation approaches, including rule-based semantic matching~\cite{meng2015rule}, collaborative filtering~\cite{wu2015collaborative}, and ontology-based matching~\cite{chen2009building}, provide important foundations for service computing. These approaches often assume a relatively static service set and lack the flexibility to handle frequent service changes. Deep learning–based approaches, such as representation learning with pre-trained encoders~\cite{wang2021servicebert}, graph neural networks~\cite{liu2023dysr}, and sequence models~\cite{fan2023service}, improve semantic understanding and recommendation accuracy. However, they typically rely on historical training data and struggle to adapt when services evolve rapidly, resulting in performance degradation and outdated recommendations. 

\begin{figure}[t]
    \centering
    \includegraphics[width=0.99\linewidth]{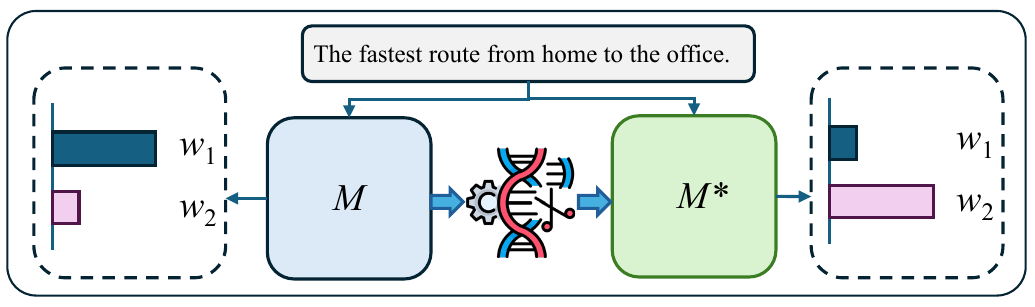}
    \caption{Model Editing for Service Evolution. $M$ denotes the original LLM, $M^*$ the updated model after service evolution, and $w_1, w_2$ represent the selection probabilities of two services, whose values change accordingly during the transition.
}\label{fig:model_editing}
\end{figure}
The emergence of large language models (LLMs) has opened new opportunities for service recommendation~\cite{ren2024representation}. LLMs possess strong abilities in semantic understanding~\cite{yang2023study}, reasoning~\cite{yao2024lawyer}, and code generation~\cite{chang2023self}, making them promising for generating executable service pipelines directly from natural language requirements. Recent advances in retrieval-augmented generation (RAG)~\cite{wu2024retrieval}, parameter-efficient fine-tuning (PEFT)~\cite{ding2023parameter}, in-context learning~\cite{dong2024survey}, and chain of thought (CoT)~\cite{wei2022chain} further enhance their adaptability across domains and tasks. \add{Moreover, LLM-based tool-use and API invocation paradigms, e.g., Toolformer~\cite{schick2023toolformer}, ReAct~\cite{yao2022react}, Gorilla~\cite{patil2024gorilla}, enable models to leverage external tools and knowledge through retrieval.} 

Despite these advances, directly applying LLMs to service recommendation still faces several challenges that limit their effectiveness in practice.
As illustrated in Fig.~\ref{fig:model_editing}, consider a service recommendation scenario where an LLM generates an executable pipeline to construct a complete travel itinerary by integrating flight booking, hotel reservation, and local transportation services. 
\add{$M$ denotes a general-purpose pretrained LLM, while $M^*$ represents the updated model after applying localized edits to incorporate newly introduced or modified service knowledge. Importantly, $M^*$ is not retrained through full model retraining, but via efficient parameter-level editing. }
When underlying services evolve, e.g., API changes or parameter updates, previously valid pipelines may become invalid, requiring model updates. However, directly updating LLMs is costly and may cause performance degradation on existing services. \add{However, injecting knowledge with a prompt at inference time without modifying the model’s internal knowledge may lead to inconsistent or unstable outputs, often requiring careful prompt engineering~\cite{weiretrieval}. Instead of full model retraining, model editing performs localized edits that modify only a small subset of parameters, which reduces the computational cost and allows efficient adaptation to evolving service ecosystems~\cite{zheng2025learning}.} Moreover, LLMs may hallucinate non-existent or outdated services, i.e., generating service names or invocations that are syntactically plausible but do not correspond to any available or executable services in the target ecosystem, resulting in pipelines that appear plausible but fail during execution~\cite{gao2025systematic}. These challenges can be summarized into the following three key aspects:

\begin{itemize}
    \item \textbf{Knowledge staleness} occurs because LLMs are trained on static data, which prevents them from naturally reflecting service updates or version changes.
    \item \textbf{Hallucinations} occur when LLMs generate non-existent or incorrect service calls, thereby undermining reliability.
    \item \textbf{Maintenance cost} arises because frequent re-training or full fine-tuning is computationally expensive and impractical for evolving ecosystems.
\end{itemize}

To address these issues, model editing has emerged as a promising paradigm that aims to modify model behavior at particular knowledge points while preserving performance on unrelated inputs. Unlike full fine-tuning, model editing enables localized modifications to model knowledge or behavior, allowing the correction or injection of specific service facts without retraining the entire model~\cite{meng2022locating, zhang2024comprehensive}. This creates a new perspective for building evolution-aware generative frameworks for service recommendation in service computing. However, existing model editing methods primarily focus on injecting or correcting isolated knowledge, while largely overlooking shifts in the underlying service data distribution \add{caused by service evolution}.
As service environments evolve, frequent and continual edits may lead to unstable model behavior and cumulative errors, undermining long-term reliability.

In this paper, we propose an \textbf{Evo}lution-Aware Framework for Service \textbf{Rec}ommendation named \tool, based on Knowledge Updating and constrained Decoding.
\tool integrates model editing and constrained decoding into a unified framework to handle dynamic service ecosystems. Specifically, model editing enables the LLM to efficiently incorporate evolving service knowledge, allowing it to adapt to interface changes and updated service facts without requiring full retraining. Constrained decoding, implemented via trie-guided  Finite Automata (FA), ensures that the generated service recommendations are executable and compliant with service interface constraints, effectively mitigating potential hallucinations. 
Extensive experiments on real-world service datasets demonstrate that our framework significantly outperforms traditional deep learning–based baselines, achieving higher recommendation accuracy and better adaptability to service evolution while maintaining low maintenance cost. 

The main contributions of this work can be summarized as follows:

\begin{itemize}
    \item To address the challenges posed by dynamic service ecosystems, we propose a unified framework that leverages model editing and trie-based constrained decoding to generate adaptive service recommendations.
    \item We adopt FA-based constrained decoding to enforce service and composition constraints, effectively mitigating hallucinations and ensuring that generated sequences are valid.
    \item We validate our approach on real-world service datasets, demonstrating \add{up to 25.9\% improvement in Recall@5} over both traditional and deep learning–based baselines in terms of recommendation accuracy and adaptability to evolving services. \add{Our code is realsed on GitHub~\footnote{https://github.com/GuodongFan/EVOREC}. }
\end{itemize}

\section{Background}

\subsection{Problem Definition}

We consider the problem of service recommendation in an evolving service ecosystem. 
Let $\mathcal{S}_t = \{s_1^t, s_2^t, \dots, s_{N_t}^t\}$ denote the set of available services at time $t$, 
where each service $s_i^t$ has an interface description, input/output constraints, and version information. 
\add{For example, in a travel planning scenario, services may include services such as ``FlightSearch'', ``FlightBooking'', and ``HotelReservation'', each requiring specific parameters, e.g., departure city, date, and producing outputs that serve as inputs to subsequent services.}
Given a user requirement expressed as a natural language query $q_t$ or code context $c_t$, the goal is to generate a sequence of services 
$Y_t = (y_1, y_2, \dots, y_T)$ such that each service in $Y_t$ complies with its interface constraints and the 
sequence adapts to the current state of the evolving service ecosystem, including new, updated, or deprecated services, without requiring full model retraining.
\add{In the above travel example, $q_t$ is a query description such as ``plan a trip from Beijing to Shanghai'', $c_t$ can provide partial code or API usage context, \add{e.g., retrieved examples as contextual demonstrations,} and $Y_t$ corresponds to a valid service sequence result.}

Formally, the task is to learn a generative function:
\begin{equation}
    f_\theta: ( q_t \lor c_t, \mathcal{S}_t) \mapsto Y_t,
\end{equation}
where $\theta$ is the model parameterized to incorporate evolving service knowledge and produce sequences that are valid, executable, and up-to-date. 
Our framework leverages model editing to handle evolving knowledge efficiently and FA-based constrained decoding to enforce interface and sequence constraints.

\subsection{\add{Brief Introduction and} Definition of Model Editing}

\add{Model editing typically follows a locate--then--edit paradigm, as exemplified by methods such as ROME~\cite{meng2022locating}. Given a user query, the model first locates the internal neurons or key–value representations that are most responsible for encoding this knowledge. Then, it performs a localized edit by modifying only a small subset of parameters to inject updated knowledge. }

Let the pre-editing model be denoted as $M$, and the post-editing model as $M^*$.  
Each editing operation modifies one knowledge point $k$, which consists of a query $x_k$ and its corresponding answer $y_k$.  
The objective of model editing can be formally expressed as follows:

\begin{equation}
M^*(x) =
\begin{cases}
y_k, & \text{if } x = x_k \text{ or } x \text{ is related to } x_k, \\
M(x), & \text{if } x \text{ is irrelevant to } x_k.
\end{cases}
\end{equation}

\add{In practice, $(x_k, y_k)$ pairs are automatically constructed from service registries, API documentation, and historical usage data, where each pair corresponds to a requirement-service pair.}

\subsection{Finite Automata and Constrained Decoding}

An FA is a computational model with a finite set of states and no auxiliary memory structure, capable of recognizing regular patterns. Formally, a deterministic finite automaton is defined as
\begin{equation}
    A = (Q, \Sigma, \delta, q_0, F),
\end{equation}
where $Q$ is a finite set of states, $\Sigma$ is the input alphabet, 
$\delta: Q \times \Sigma \rightarrow Q$ is the transition function, 
$q_0 \in Q$ is the start state, and $F \subseteq Q$ is the set of accepting states. 
Compared to pushdown automata, finite automata lack a stack and therefore capture only linear, non-nested structural constraints, but are simpler and more efficient to apply in decoding.

During constrained decoding, the finite automaton regulates the output of a language model by enforcing regular-pattern constraints. At each generation step, a token $x_t$ is permitted only if a valid automaton transition
\begin{equation}
    q_t = \delta(q_{t-1}, x_t)
\end{equation}
exists, where $q_{t-1}$ is the current state and $q_t$ is the next state after consuming the token. Tokens that do not correspond to any valid transition are pruned during decoding, ensuring that the generated sequence conforms to the prescribed regular structure, such as API invocation formats, ordered protocol fields, or fixed template patterns.

Unlike a Pushdown Automaton, an FA does not possess explicit memory in the classical computational sense. The decoding process of the FA allows it to behave as if previously generated content were effectively remembered. After producing a prefix $x_1, \ldots, x_{t-1}$, the automaton does not store this prefix directly; rather, its effect is fully encoded by the current state $q_{t-1}$. This state represents an equivalence class of all prefixes that admit the same set of valid future continuations. In this sense, the generated prefix functions as an external form of 
memory: the automaton restricts future tokens based on the state reached through processing the prefix. Consequently, even though finite automata lack auxiliary memory structures such as stacks, their integration with the autoregressive generation process enables them to utilize past information implicitly through the 
state transitions.

\subsection{Service Recommendation from Natural Language Requirements or Code Context}

Service recommendation aims to map a user \add{query} \( q_t \), typically expressed in natural language, or a code context \(c_t\), e.g., a code snippet, to a set of candidate services \( \mathcal{S} = \{s_1, s_2, \dots, s_n\} \). Formally, the recommendation task can be expressed as:

\begin{equation}
\hat{\mathcal{S}}_k
=
\underset{\substack{\mathcal{S}' \subseteq \mathcal{S}}}{\arg\max}
\sum_{s_i \in \mathcal{S}'} \operatorname{sim}(q_t/c_t, s_i).
\end{equation}
where $|\mathcal{S}'| = k$ and \(\text{sim}(\cdot, \cdot)\) measures the semantic similarity between the query/code context and each candidate service, e.g., using embeddings or textual matching. The goal is to \add{generate service recommendations}
\(\hat{\mathcal{S}}_k \subseteq \mathcal{S}\)
that best satisfies the user's requirements.

\section{Approach}

\begin{figure*}[t]
\centering
\includegraphics[width=0.9\linewidth]{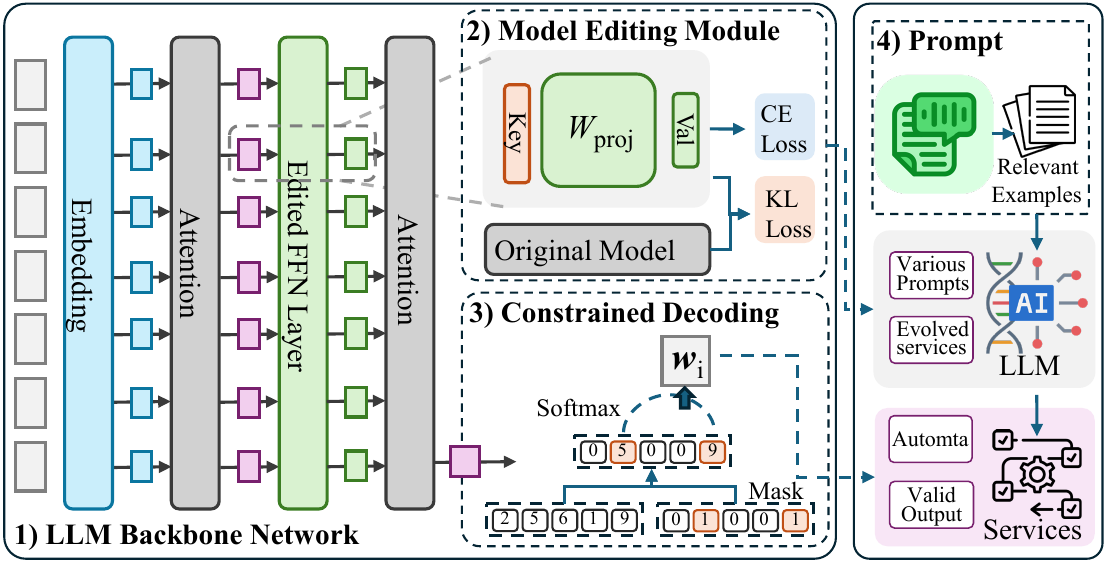}
\caption{Overall framework of the proposed \tool. Model editing incorporates evolving service knowledge into the LLM, while FA-based constrained decoding ensures the validation of the generated sequence.}\label{fig:framework}
\end{figure*}

\add{The constrained decoding mechanism reduces hallucinations, i.e., preventing invalid, non-existent or incompatible services, by restricting the generation space to valid service names and enforcing structural constraints. While model editing helps reduce distribution mismatch by aligning the model’s internal knowledge with updated service, i.e., adding and updating services, thereby improving consistency in downstream generation. The input query is augmented via retrieval, processed by the LLM, adapted through model editing, and finally decoded under FA-based constraints to generate valid service sequences.} As illustrated in Fig.~\ref{fig:framework}, our approach consists of four main components: 

\begin{itemize}
    \item \textbf{LLM Backbone Network.} A Transformer-based decoder that encodes natural language requirements or code context into related services.
    \item \textbf{Model Editing Module.} An edited FFN layer with key-value projections adapts the model to evolving knowledge. It is optimized with cross-entropy loss for correctness and KL-divergence loss for consistency. \add{The edits generalize to semantically similar queries while preserving unrelated service knowledge.}
    \item \textbf{Constrained Decoding.} A decoding strategy that filters and re-weights candidate tokens to enforce validity and reduce invalid outputs.
    \item \textbf{Retrieval-Augmented Prompt.} 
    A retrieval module dynamically supplies relevant examples from a service corpus to construct augmented prompts, enhancing the model’s contextual understanding and adaptability to new service knowledge.
\end{itemize}


\subsection{Model Editing for Knowledge Updating}

We adopt a locate-then-edit approach, ROME, for knowledge updating~\cite{meng2022locating}. \add{Intuitively, this process consists of two steps: locating where the target service knowledge is stored (key), and updating the corresponding representation (value) to reflect new or modified services. In essence, the editing process injects a mapping from a requirement pattern to the target services into the model’s internal space.}

\subsubsection{Calculate Key Value}
The key vector represents the internal activation pattern that identifies where the target knowledge is stored in the model, as \add{shown in Eq.~\ref{eq:edit_key}. In the service recommendation setting, it corresponds to the representation of a service requirement, e.g., a query or code context, that is associated with a specific service.}

\begin{equation}\label{eq:edit_key}
k^{\ast} = \frac{1}{N} \sum_{j=1}^{N} k(x_j + s),
\end{equation}
where \(x_j\) denotes the \(j\)-th randomly sampled prefix and \(k(x_j + s)\) is the activation of the final token of \(s\) at the input of
the feed-forward layer being edited (i.e., the input to \(W_{\text{proj}}\)).


\subsubsection{Update Value}
The value vector \(v\) encodes the updated factual content to be written into the model at the located key position. \add{It is learned via a combination of cross-entropy and KL divergence losses, as shown in Eqs.~\ref{eq:edit_value}--\ref{eq:edit_loss2}. The value vector $v$ encodes the representation of the target service to be generated, enabling the model to produce the correct service when encountering similar requirement patterns. The update is applied to the output projection matrix  $W_{\text{out}}$ of the MLP layer in a specific Transformer block. 
}
\begin{equation}\label{eq:edit_value}
\mathcal{L}(v) = \mathcal{L}_1(v) + \mathcal{L}_2(v)
\end{equation}

\begin{equation}\label{eq:edit_loss1}
\mathcal{L}_1(v) = \frac{1}{N}\sum_{j=1}^N -\log P_{M'}(s_j \mid q + p),
\end{equation}

\begin{equation}\label{eq:edit_loss2}
\mathcal{L}_2(v) = D_{KL}\!\big(P_{M'}(r \mid q') \,\|\, P_{M}(r \mid q')\big).
\end{equation}

Here, \(M\) and \(M'\) denote the original and edited models, respectively.
The query \(q\) represents a service requirement with contextual prompt \(p\),
and \(s_j\) is the corresponding target service.
The auxiliary query \(q'\) with irrelevant validation input \(r\) is used to
regularize the edit.
The loss \(\mathcal{L}_1\) enforces the edited model to produce the target service,
while \(\mathcal{L}_2\) preserves the model’s behavior on unrelated inputs via KL divergence, preventing catastrophic forgetting.

\subsubsection{Editing Knowledge on Linear Projection Layer}

Specifically, in the service recommendation setting, we regard $W_{\text{proj}}$ as a linear key-value memory, i.e., $W K \approx V$, 
where the key set $K = \{k_1, k_2, \ldots, k_n\}$ corresponds to service-related queries and the value set $V = \{v_1, v_2, \ldots, v_n\}$ corresponds to the recommended services. 
The goal is to add a new key-value pair $(k^\ast, v^\ast)$ into $W_{\text{proj}}$, such that the new service requirement $k^\ast$ is mapped to the desired service $v^\ast$, while preserving the original mappings as much as possible. 
This can be formulated as the following constrained least-squares problem, as follows. 

\begin{equation}
\min \ \lVert W K - V \rVert
\end{equation}

\begin{equation}
\text{s.t. } \ W k^\ast = v^\ast .
\end{equation}

The closed-form solution to this optimization is given by \add{Eq.~\ref{eq:Linear_w}, which performs a rank-one update to the weight matrix $W$, injecting a mapping from a requirement $k^*$ to the target service $v^*$. This localized update avoids full model retraining and restricts the modification to a specific subspace of the parameter matrix. }

\begin{equation}\label{eq:Linear_w}
\hat{W} = W + \Lambda (C^{-1} k^\ast)^{\top},
\end{equation}
where $\Lambda = (v^\ast - W k^\ast)/(C^{-1} k^\ast)^{\top} k^\ast$,  
$W$ is the original weight matrix, $\hat{W}$ is the updated weight matrix, and $C = K K^{\top}$ is a pre-computed covariance matrix of keys. 
This formulation enables efficient updates through low-rank matrix modifications. 
By leveraging this lightweight update mechanism, ROME directly inserts the mapping from the service requirement $k^\ast$ to the target service $v^\ast$, 
thus achieving knowledge editing tailored for service recommendation.

\subsection{Trie-Guided and FA-Constrained Decoding}

\add{The key idea is to restrict the model’s output space to valid service sequences by enforcing lexical and structural constraints during decoding. These constraints are derived from service interface specifications, API documentation, or service catalogs, which define valid service names, parameters, and invocation structures. Based on these sources, we construct a service lexicon and corresponding structural constraints, which are encoded into a Trie and an FA. 
The Trie ensures lexical validity of service names under subword tokenization, preventing invalid token combinations, while a FA enforces valid structural transitions during generation. }

Unconstrained decoding often yields invalid or duplicate service sequences, as shown in Alg.~\ref{alg:pda-trie-main} and Alg.~\ref{alg:allowed}. We enforce correctness with a constrained decoder that combines an FA with a service lexicon Trie. The FA governs list structure, i.e., start, in-service, separator, end, while the Trie constrains token choices to valid service names and enables service-level deduplication.

\paragraph{States and transitions}
Let $T$ be the Trie over tokenized service names. Decoding alternates between two observable phases: 
(i) \textsc{START}, where a new service may start; and 
(ii) \textsc{IN\_SERVICE}, where tokens must follow a path in $T$. 
A service is completed only when the current Trie node is terminal, and the decoder emits a comma or $\langle eos\rangle$; the corresponding service is then added to the used set $D$.

\paragraph{Allowed set}
Let $u_t$ be the Trie node after the last separator and define
$R(u) \coloneqq \{\, w \in \Lab(u) \mid \subsids(\Next(u,w)) \setminus D \neq \varnothing \,\}$. 

\begin{equation}
A_t =
\begin{cases}
  R(root)\cup\{\langle eos\rangle\}, & u_t=root,\\[2pt]
  R(u_t)\cup\{c,\langle eos\rangle\}, & \text{$u_t$ terminal and }\sid(u_t)\notin D,\\[2pt]
  R(u_t), & \text{otherwise.}
\end{cases}
\end{equation}

\paragraph{Masked decoding rule}
At step $t$, we mask the model distribution by $A_t$:
\begin{equation}
P'(w_t \mid w_{<t}) \;=\;
\begin{cases}
P(w_t \mid w_{<t}) & \text{if } w_t \in A_t,\\
0 & \text{otherwise}.
\end{cases}
\end{equation}

When $u_t$ is terminal and $w_t\in\{c,\langle eos\rangle\}$, we mark the corresponding service as used: $D \leftarrow D \cup \{sid(u_t)\}$ and (if $w_t{=}c$) reset $u_{t+1}\!=\!root$ to start the next service. This rule allows us to prevent duplicates while allowing shared prefixes, e.g., \texttt{google-}\emph{maps} vs.\ \texttt{google-}\emph{api}.

We mask model logits to permit only $A_t$. When a terminal node is closed by emitting $c$ or $\langle eos\rangle$, its $\sid(u_t)$ is inserted into $D$ and the Trie position resets to $\mathsf{root}$. This enforces deduplication while preserving shared-prefix paths toward still-unused services.


\paragraph{Remarks}
(1) Parsing the previously generated prefix no longer requires a stack; a finite automaton is sufficient during sequence generation.
(2) The key to safe dedup is subtree reachability: we block only branches whose subtrees contain no unused services, rather than banning tokens that appear in already used names.

\begin{algorithm}[t]
\caption{Trie and FA Decoding with De-duplication}
\label{alg:pda-trie-main}
\small
\textbf{Input:} LLM $\mathcal{M}$; service Trie $T$ (node $u$: $children$, $terminal\_sid$, $subtree\_sids$);
comma $c$; end $\langle eos\rangle$.\\
\textbf{Output:} valid, deduplicated sequence $\mathcal{I}$.
\begin{algorithmic}[1]
\State $\mathcal{I}\gets\epsilon$;\; $D\gets\emptyset$;\; $u\gets root(T)$ \Comment{initialize an empty output sequence, set, node}
\While{true}
  \State $\ell \gets \log\mathrm{softmax}(\mathcal{M}(\text{prefix}=\mathcal{I}))$
  \State $A \gets \textsc{Allowed}(u,D,T,c,\langle eos\rangle)$
  \State \textbf{mask:} $\ell[w]\leftarrow-\infty$ for $w\notin A$
  \State $w \gets \arg\max_{v\in A} \ell[v]$
  \State $\mathcal{I}\gets \mathcal{I}\circ w$
  \If{$w\in\{c,\langle eos\rangle\}$}
     \If{$terminal\_sid(u)\neq\bot$ \textbf{and} $terminal\_sid(u)\notin D$}
        \State $D\gets D\cup\{terminal\_sid(u)\}$
     \EndIf
  \EndIf
  \If{$w=\langle eos\rangle$}
     \textbf{break}
  \ElsIf{$w=c$}
     \State $u\gets root(T)$
  \ElsIf{$w\in children(u)$}
     \State $u\gets child(u,w)$
  \EndIf
\EndWhile
\State \Return $\mathcal{I}$

\end{algorithmic}
\end{algorithm}

\begin{algorithm}[t]
\caption{Allowed($u,D,T,c,\langle eos\rangle$): Reachable Token Set}
\label{alg:allowed}
\small
\textbf{Input:} Trie node $u$; used services $D$; Trie $T$; $c$, $\langle eos\rangle$. \\
\textbf{Output:} allowed token set $A$.
\begin{algorithmic}[1]
\State $A\gets\emptyset$;\; define $U(v)=subtree\_sids(v)\setminus D$
\If{$u = root(T)$}
   \State $A \gets \{\,w\in children(root)\mid U(child(root,w))\neq\emptyset\,\} \cup \{\langle eos\rangle\}$
\Else
   \State $A \gets \{\,w\in children(u)\mid U(child(u,w))\neq\emptyset\,\}$
   \If{$terminal\_sid(u)\notin D$}
      \State $A \gets A \cup \{c,\langle eos\rangle\}$
   \EndIf
\EndIf
\State \Return $A$
\end{algorithmic}
\end{algorithm}


\begin{figure*}[t]
\centering
\begin{subfigure}[t]{0.49\textwidth}  
\centering
\begin{tikzpicture}[>=Stealth,font=\scriptsize,node distance=7mm and 9mm,
  blk/.style={draw,rounded corners,inner sep=2pt,minimum height=4.6mm,align=center},
  state/.style={blk,minimum width=19mm},
  allow/.style={draw=black,line width=0.8pt}]
\node[state] (S0) {START};
\node[state, right=9mm of S0] (S1) {IN\_SERVICE};
\node[state, right=9mm of S1] (S2) {CLOSE};
\draw[->,allow] (S0) -- node[above=0mm]{start} (S1);
\draw[->,allow] (S1) -- node[below=0.2mm]{$c$ / $\langle eos\rangle$} (S2);
\draw[->,allow] ([yshift=1mm]S2.north east) to[out=80,in=80,looseness=0.55]
  node[above=0mm]{comma} ([yshift=1mm]S0.north west);
\node[blk, below=8mm of S1, minimum width=33mm] (AT) {$A_t=\mathrm{Allowed}(u_t,D)$};
\draw[->,allow] (S1) -- (AT);
\node[blk, right=8mm of AT, minimum width=26mm] (MASK) {mask logits by $A_t$};
\draw[->,allow] (AT) -- (MASK);
\end{tikzpicture}
\caption{FA flow and masking by $A_t$. The decoding process is governed by an implicit DFA induced by the service Trie and token-level control flow.}
\end{subfigure}\hfill%
\begin{subfigure}[t]{0.49\textwidth}  %
\resizebox{0.7\linewidth}{!}{%
\centering
\begin{tikzpicture}[>=Stealth,font=\scriptsize,node distance=6.5mm and 8mm, 
  dot/.style={circle,draw,minimum size=4.2mm,inner sep=0pt},
  utnode/.style={circle,draw,double,double distance=1pt,minimum size=4.4mm,inner sep=0pt}, 
  allow/.style={draw=black,line width=0.8pt},
  prune/.style={draw=black!65,densely dashed,line width=0.7pt},
  box/.style={draw,rounded corners,inner sep=2pt,align=center}]
\node[dot] (root) {};
\node[anchor=south west] at ($(root)+(-2.5mm,2.6mm)$) {root};

\node[dot, right=7.5mm of root] (a1) {};
\node[dot, right=7.0mm of a1] (a2) {};
\node[dot, right=7.0mm of a2] (a3) {};
\draw[->,allow] (root) -- node[above=-1pt] {$a$} (a1);
\draw[->,allow] (a1) -- node[above=-1pt] {$b$} (a2);
\draw[->,allow] (a2) -- node[above=-1pt] {$c$} (a3);
\node[anchor=west] at ($(a3.south east)+(-12mm,-1.8mm)$) {\tiny terminal, $\mathrm{sid}=7$};

\node[utnode, below=7.5mm of a1] (x1) {};           
\draw[->,allow] (root) -- node[left=-1pt] {$x$} (x1);
\node[anchor=west] at ($(x1.south east)+(-12mm,-1.8mm)$) {\tiny terminal, $\mathrm{sid}=3$ (prefix)};
\node[dot, right=7.0mm of x1] (x2) {};
\node[dot, right=7.0mm of x2] (x3) {};
\draw[->,allow] (x1) -- node[above=-1pt] {$y$} (x2);
\draw[->,allow] (x2) -- node[above=-1pt] {$z$} (x3);
\node[anchor=west] at ($(x3.south east)+(-12mm,-1.8mm)$) {\tiny terminal, $\mathrm{sid}=12$};

\node[dot, below=8mm of root] (g1) {};
\draw[->,prune] (root) -- node[left=-1pt] {$g$} (g1);
\draw[->,prune] (g1) -- node[above=-1pt] {$o$} (x1); 

\node[box, below=7mm of x2, minimum width=42mm] (D)
  {$D=\{\mathrm{sid}=3,\ \mathrm{sid}=5,\dots\}$\\ \tiny pruned subtrees only in $D$};
\end{tikzpicture}
}
\caption{Trie reachability, where subtrees are pruned only when all reachable service identifiers are contained in $D$.}
\end{subfigure}
\caption{Trie-guided FA constrained decoding.}
\label{fig:trie-guided-subfigs}
\end{figure*}
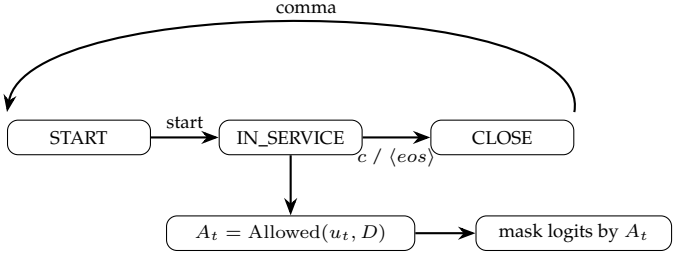
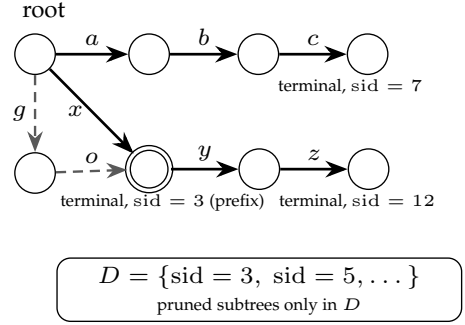

\subsection{Retrieval-Augmented Prompt Design}
\label{sec:prompt}

While model editing allows incorporating new knowledge into LLMs, injecting too much knowledge can cause degradation\cite{zhang2024comprehensive}. 
Therefore, we integrate a retrieval module as an alternative to dynamically provide relevant knowledge with prompt-based generation. 

\paragraph{Retrieval Augmentation}
Given a query description $q$, the model retrieves semantically 
similar examples from a corpus 
$\mathcal{C}=\{(x_i, y_i)\}_{i=1}^{N}$, 
where $x_i$ is an application description and $y_i$ the associated APIs. 
Each description is encoded into a dense embedding 
$\mathbf{v}_i=f_\theta(x_i)$ using a sentence-transformer encoder, ``all-MiniLM-L6-v2''. 
\add{The corpus $C$ is constructed from the training datasets, and in practice, it can be the latest API repository. }
A cross-encoder reranker, ``bge-reranker-base'' is then applied to re-rank the initial candidates for improved relevance.
The query is similarly encoded as $\mathbf{v}_q=f_\theta(q)$.  
We compute cosine similarity as:

\begin{equation}
\mathrm{sim}(\mathbf{v}_q,\mathbf{v}_i)
=\frac{\mathbf{v}_q^\top \mathbf{v}_i}{\|\mathbf{v}_q\|\|\mathbf{v}_i\|},
\quad
\mathcal{N}_k(q)=\mathrm{arg\,topk}_{i\in\mathcal{C}}\mathrm{sim}(\mathbf{v}_q,\mathbf{v}_i).
\end{equation}

The top-$k$ examples $\mathcal{N}_k(q)$ are retrieved to serve as few-shot contextual demonstrations.  
During training, the identical description is removed to avoid self-matching.

\paragraph{Prompt Construction}
The final prompt follows a structure consisting of an instruction, an input, and a set of few-shot examples. 
It begins with a task instruction that defines the model's role as an API recommendation expert, 
followed by the input description and category information.  
The retrieved examples $\mathcal{N}_k(q)$ are then appended as the Similar Examples, each including a short description, its categories, and related APIs.  
This retrieval-augmented design allows the LLM to reason from semantically related cases while maintaining a consistent, structured output. 
An abstracted version of the Prompt Template is shown in Fig.~\ref{fig:prompt-template}.

\begin{figure}[h]
\centering
\small
\begin{tcolorbox}[
    colback=promptgray,
    colframe=gray!30,
    boxrule=0.3pt,
    arc=3pt,
    left=6pt,
    right=6pt,
    top=5pt,
    bottom=5pt,
    width=0.92\linewidth,
    fonttitle=\bfseries,
]
\ttfamily
\textcolor{promptblue}{You are an expert in {} API and service recommendation for mashup applications.}\\
\textcolor{promptblue}{Your task is to read a short description of an application idea and output a list of the most relevant APIs that can support its implementation.}\\[4pt]

\textcolor{promptpurple}{Input description:} \textless description\textgreater\\
\textcolor{promptpurple}{Categories:} \textless categories\textgreater\\[3pt]

\textcolor{promptpurple}{Similar examples:}\\
\ \ 1. \textcolor{promptgreen}{Description:} \textless desc\textgreater\\
\ \ \ \ \textcolor{promptgreen}{Categories:} \textless categories\textgreater\\
\ \ \ \ \textcolor{promptgreen}{Related APIs:} \textless apis\textgreater\\[3pt]

\textcolor{promptpurple}{Output:}
\end{tcolorbox}
\caption{Prompt template used for service recommendations.}
\label{fig:prompt-template}
\end{figure}

This unified retrieval-augmented prompt enables the model to leverage 
semantic similarity for in-context reasoning and produces more relevant API recommendations, \add{ serving as a knowledge supplement for model editing.}

\section{Experimental Setup}

\subsection{Dataset}

\begin{figure}[t]
\centering
\begin{tikzpicture}[
    font=\small,
    >=Latex,
    box/.style={draw, rounded corners, minimum width=1.35cm, minimum height=0.55cm, align=center},
    train/.style={box, fill=blue!12},
    test/.style={box, fill=green!15},
    seglabel/.style={font=\bfseries\footnotesize}
]

\draw[->, thick] (0,0) -- (6.6,0);

\draw[dashed] (2.2,0.3) -- (2.2,-2.0);
\draw[dashed] (4.4,0.3) -- (4.4,-2.0);

\node[seglabel] at (1.1,-0.25) {$t_1$};
\node[seglabel] at (3.3,-0.25) {$t_2$};
\node[seglabel] at (5.5,-0.25) {$t_3$};

\node[train] at (1.1,-0.95) {Train};
\node[test]  at (1.1,-1.55) {Test};

\node[train] at (3.3,-0.95) {Train};
\node[test]  at (3.3,-1.55) {Test};

\node[train] at (5.5,-0.95) {Train};
\node[test]  at (5.5,-1.55) {Test};

\end{tikzpicture}
\caption{Chronological Dataset Split.}
\label{fig:time_split}
\end{figure}
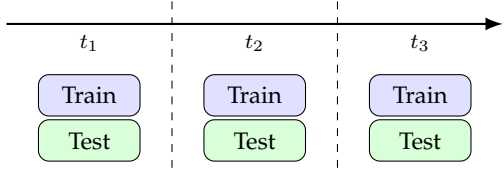

We train our model with web service data crawled from ProgrammableWeb, the largest online API registry, which contains 22,457 Web APIs and 8,217 mashups. In addition, we conduct comparative experiments on another dataset containing \add{3,925} Python and \add{6,563} Java APIs along with corresponding queries without using the augmented code data\add{, adopted from our prior work~\cite{fan2023service}}. \add{We split the dataset into 70\% for training and 30\% for testing.} Each data point in this dataset is represented as 
$\langle Requirements, \{API_1, API_2, ..., API_n\} \rangle$, 
where $n$ denotes the number of candidate APIs for each query. The user requirements within the dataset pertain to either the description of mashups or functional specifications related to specific programming languages~\cite{fan2023service}. \add{Moreover, we construct an evolution-oriented split by ordering all mashups according to their release date in ascending order, and then dividing the dataset chronologically into three consecutive segments, as shown in Fig.~\ref{fig:time_split}. For each segment, we further split the data into training and test sets, and the evaluation is performed on later instances, better reflecting real-world deployment under the evolution of the service ecosystem.}

To comprehensively assess the effectiveness and generalization of our approach, we further include a code-based dataset that provides real programming contexts for API invocation~\cite{peng2023}. Specifically, the dataset consists of code snippets where the surrounding context, such as function definitions, variable names, and comments, serves as input, and the target is to predict the correct API to fill in. This dataset enables our model to learn contextual code-level semantics and improve API recommendation accuracy in practical programming scenarios.

We observe an inconsistency between training and constrained decoding. In this setting, subword tokenizers often merge delimiters, e.g., ``(),'', into a single token, while the FA decoder assumes explicit token boundaries.
This mismatch prevents \tool from generating multiple API names, as the model does not learn to produce a standalone comma token, causing the FA to terminate after a single API name.
We address this by replacing commas with a dedicated single-token separator $<|$SEP$|>$, ensuring consistent segmentation and stable constrained generation.

\subsection{Model Parameters}

Table~\ref{tab:model-params} summarizes the main hyperparameters used in our experiments. 
The Qwen2.5 model serves as the base architecture, trained with mixed-precision computation, and updated through a single editable layer located at the fifth layer. \add{
We adopt a backbone model with 7B parameters, which represents a commonly used scale for practical deployment while maintaining a balance between performance and computational cost~\cite{jiang2025aixcoder}.}
For optimization, we set the number of gradient steps to 40, the learning rate to 1e-2, and apply a weight decay of 1e-3. 
To control parameter updates, the clamp factor is fixed at 4, and the KL regularization factor is set to 0.06. 

Regarding retrieval configuration, we use all-MiniLM-L6-v2 as the retrieval model, with a retrieval count of 5. 
The context template length is set to [5,8] [8,8], and model parallelism is enabled (true) to support efficient computation. 
The language modeling head (lm\_head) remains consistent with the default architecture of the base model. \add{Unless otherwise specified, all experiments are conducted in accordance with the setup described in this section.}

\begin{table}[h]
\centering
\caption{Model Parameters.}
\small
\setlength{\tabcolsep}{1pt}
\renewcommand{\arraystretch}{1.1}
\begin{tabularx}{\linewidth}{lclc}
\toprule
\textbf{Parameter} & \textbf{Value} & \textbf{Parameter} & \textbf{Value} \\
\midrule
fact token & \texttt{last subject} & layers & \texttt{5}  \\
num grad steps & \texttt{40} & v lr & \texttt{1e-2} \\
loss layer & \texttt{27} & weight decay & \texttt{1e-3} \\
clamp factor & \texttt{4} & kl factor & \texttt{0.06} \\
context len & \texttt{$[5,8] [8,8]$} & model parallel & \texttt{true} \\
fp16 & \texttt{true} & model name & \texttt{Qwen2.5} \\
lm head & \texttt{lm\_head} & retrieval count & \texttt{5} \\
retrieval model & \texttt{all-MiniLM-L6-v2} \\
\bottomrule
\end{tabularx}\label{tab:model-params}
\end{table}

\subsection{Evaluation Metrics} 
To comprehensively evaluate the performance of our model, we adopt several widely used ranking metrics, including Recall@K, Precision@K, and Mean Average Precision (mAP@K).

The Recall@K and Precision@K metrics are defined as follows:

{\small
\begin{equation}
\text{Recall@}K = \frac{TP}{TP + FN}, \quad 
\text{Precision@}K = \frac{TP}{TP + FP},
\end{equation}
}
where $TP$, $FP$, and $FN$ denote the number of true positives, false positives, and false negatives, respectively.

To further measure the ranking quality, we use the mAP@K, which means value across all test samples, as follows:

{\small
\begin{equation}
AP@K = \frac{1}{m} \sum_{k=1}^{K} P(k) \cdot rel(k), \quad 
mAP@K = \frac{1}{|M|} \sum_{m=1}^{|M|} (AP@K)_m,
\end{equation}
}
where $P(k)$ denotes the precision at cutoff $k$, $rel(k)$ is an indicator function that equals 1 if the $k$-th item is relevant and 0 otherwise, and $|M|$ is the total number of test mashups.

These metrics jointly evaluate both retrieval accuracy and the relative ranking quality of the recommended results. 

\subsection{Research Questions}

To comprehensively evaluate the effectiveness of our approach, we formulate the following research questions (RQs):

\textbf{RQ1:} How does the proposed method perform compared to baseline approaches?
We answer this RQ by comparing our approach with representative baselines across multiple datasets using standard evaluation metrics. 
\add{This RQ also reflects the effectiveness of implicit knowledge integration, e.g., model editing, compared to methods without explicit knowledge updating.}

\textbf{RQ2:} How does each component of the proposed framework contribute to the overall performance?
This RQ is investigated through systematic ablation studies that remove or modify individual components of the framework.

\textbf{RQ3:} How effective is the proposed approach in handling software evolution scenarios?
We address this RQ by evaluating the model on real-world software evolution cases, such as API changes across versions, and measuring its ability to correctly adapt or edit code accordingly.

\textbf{RQ4:} How well does the proposed method support recommendations in real-world code environments?
This RQ is examined by deploying our approach in practical code settings and assessing its recommendation performance under realistic constraints.

\textbf{RQ5:} What is the impact of constrained decoding on the model’s probability distribution?
This RQ is investigated by analyzing probability cost, entropy changes, and validity–probability tradeoffs, to understand how constraints affect model confidence.

\subsection{Experimental Environment}
All experiments are conducted on a machine equipped with an NVIDIA GeForce RTX 5090 GPU, an AMD Ryzen 9 9950 CPU, and 128\,GB of system memory. Experiments are performed on 3B- and 7B-parameter models. We conduct our experiments using the EasyEdit framework\footnote{https://github.com/zjunlp/EasyEdit}.

\subsection{Baseline Selection}

To comprehensively evaluate the effectiveness of our proposed approach, we compare it against a diverse set of representative baselines that cover different modeling paradigms for API/mashup recommendation tasks.

Specifically, we include traditional information retrieval (IR)-based methods, which rely on textual similarity between mashup descriptions and mashup-related information in the dataset and serve as strong non-neural baselines. In addition, we consider several neural and representation-learning-based approaches, including MTFM++~\cite{wu2021mashup} and MNT, which model semantic relationships between mashups and APIs using learned latent representations.
We further incorporate GSR~\cite{fan2023service}, a generative semantic representation model designed to capture higher-level semantic information. We adopt a fine-tuning-based approach using LLamaFactory~\cite{zheng2024llamafactory} as an upper bound, which adapts pre-trained models to the target task. 
\add{In addition, we include a RAG baseline implemented with a retrieval count of 5, representing inference-time knowledge integration without modifying model parameters. We further adopt an agent-based baseline following a ReAct-style paradigm, where the model performs iterative reasoning and tool interaction by retrieving relevant examples and refining its decisions over multiple steps~\cite{yao2022react}. For a fair comparison, we adopt the base model ``Qwen2.5-7B-Instruct'' and the retrieval model ``all-MiniLML6-v2''. }


\section{Experimental Results}

In this section, we systematically address the four research questions by demonstrating the overall effectiveness of the proposed method, its superiority over both traditional and deep-learning-based baselines, the contributions of key components and design choices, as well as its robustness and generalization across diverse datasets and experimental settings.

\subsection{RQ1. Baseline Comparison}

\begin{table*}[t]
\centering
\caption{Performance Comparison. FT represents a fine-tuning baseline, where the entire model is retrained using all available data, serving as an upper-bound adaptation strategy.}
\label{tab:perf}
\resizebox{\textwidth}{!}{
\begin{tabular}{l l c c c c c c c c c}
\toprule
Dataset & Model & Recall@1 & Precision@1 & mAP@1 & Recall@5 & Precision@5 & mAP@5 & Recall@10 & Precision@10 & mAP@10\\
\midrule
\multirow{8}{*}{Mashup}
& IR        & 0.328 & 0.469 & 0.469 & 0.419 & \second{0.427} & 0.520 & 0.423 & \second{0.426} & 0.520 \\
& MTFM++    & 0.401 & 0.579 & 0.579 & 0.563 & 0.187 & 0.625 & 0.624 & 0.108 & 0.618 \\
& MNT       & 0.452 & 0.643 & 0.643 & 0.564 & 0.197 & 0.673 & 0.567 & 0.159 & 0.672 \\
& GSR GEN              & 0.491 & \second{0.695} & \second{0.695} & 0.549 & \best{0.643} & 0.710 & 0.549 & \best{0.643} & \second{0.710} \\
\cdashline{2-11}
& RAG 
& 0.225 & 0.295 & 0.295 
& 0.362 & 0.251 & 0.393 
& 0.368 & 0.247 & 0.394 \\ 
& Agent 
& 0.485 & 0.680 & 0.680 
& 0.617 & 0.500 & 0.721 
& 0.625 & 0.495 & 0.720 \\
& FT              & 0.520  & 0.732  & 0.732  & 0.617  &  0.669 &  0.701 & 0.620  & 0.664  & 0.700  \\
\cmidrule(lr){2-11}
& \tool & \best{0.514} & \best{0.724} & \best{0.724} & \best{0.685} &  0.233 & \best{0.760} &  \best{0.685} &  0.233 & \best{0.760}\\ 

\midrule
\multirow{8}{*}{Java}
& IR        & 0.183 & 0.300 & 0.300 & 0.289 & 0.287 & 0.328 & 0.289 & 0.287 & 0.328 \\
& MNT        & 0.166 & 0.269 & 0.269 & 0.307 & 0.157 & 0.315 & 0.317 & 0.153 & 0.316 \\
& MTFM++\cite{wu2021mashup} & 0.036 & 0.060 & 0.060 & 0.109 & 0.039 & 0.100 & 0.158 & 0.029 & 0.103 \\
& GSR GEN              & \second{0.196} & \second{0.316} & \second{0.316} & \second{0.343} & \second{0.205} & \second{0.360} & \second{0.349} & \second{0.203} & \second{0.360} \\
\cdashline{2-11}
& RAG 
& 0.121 & 0.188 & 0.188 
& 0.211 & 0.210 & 0.245 
& 0.212 & 0.210 & 0.245 \\
& Agent 
& 0.221 & 0.333 & 0.333 
& 0.335 & 0.266 & 0.377 
& 0.360 & 0.260 & 0.379 \\
& FT              & 0.234  & 0.367  &  0.367 &  0.330 & 0.357  & 0.367  & 0.330  & 0.357 & 0.367  \\
\cmidrule(lr){2-11}
& \tool & \best{0.221} & \best{0.339} & \best{0.344} & \best{0.382} & \best{0.205} & \best{0.401} & \best{0.382} & \best{0.205} & \best{0.401} \\
\midrule
\multirow{7}{*}{Python}
& IR        & 0.082 & 0.159 & 0.159 & 0.154 & 0.150 & 0.201 & 0.156 & 0.151 & 0.202 \\
& MNT       & 0.118 & 0.234 & 0.234 & 0.250 & 0.147 & 0.291 & 0.252 & 0.143 & 0.291 \\
& MTFM++\cite{wu2021mashup} & 0.100 & 0.203 & 0.203 & 0.249 & 0.106 & 0.268 & 0.312 & 0.068 & 0.265\\
& GSR GEN              & \second{0.122} & \second{0.238} & \second{0.238} & \second{0.252} & \second{0.155} & \second{0.297} & \second{0.254} & \second{0.152} & \second{0.296} \\
\cdashline{2-11}
& RAG 
& 0.089 & 0.166 & 0.166 
& 0.183 & 0.181 & 0.233 
& 0.183 & 0.180 & 0.233 \\
& Agent 
& 0.146 & 0.261 & 0.261 
& 0.249 & 0.201 & 0.306 
& 0.273 & 0.193 & 0.307 \\
& FT              &  0.187  & 0.335 & 0.335  & 0.287 & 0.284 & 0.292  &  0.287 & 0.284 & 0.292 \\
\cmidrule(lr){2-11}
& \tool & \best{0.189} & \best{0.312} &  \best{0.312} & \best{0.373} & \best{0.197} &  \best{0.402} & \best{0.373} & \best{0.197} & \best{0.402} \\
\bottomrule
\end{tabular}}
\end{table*}

Table~\ref{tab:perf} presents a comprehensive performance comparison across all datasets. Several observations can be drawn from the results. 
First, traditional IR-based and early neural models, i.e., IR, MTFM++, and MNT, exhibit limited robustness. Their performance fluctuates substantially across datasets, indicating they are unable to capture service semantics. For example, on the Mashup dataset, IR reaches only 0.328 Recall@1, while on Java and Python, this value drops to 0.183 and 0.082, respectively. MTFM++ suffers even more severe degradation, achieving only 0.036 Recall@1 on Java, highlighting its sensitivity to distributional differences across datasets.
Second, GSR~GEN provides a stronger content-aware baseline and shows noticeable improvements over earlier models. Nevertheless, its advantages diminish at larger retrieval depths. On Mashup, GSR~GEN achieves Recall@1 of 0.491, but Recall@10 remains at 0.549, lagging behind FT and \tool. This pattern suggests that a generative model alone is insufficient for capturing fine-grained semantic patterns.
Third, FT yields substantial performance gains by adapting the model to updated service information. Taking Mashup as an example, FT improves Recall@1 from GSR~GEN's 0.491 to 0.520, representing a relative gain of 5.9\%. On Python, FT boosts Recall@1 from 0.122 to 0.187, a relative increase of 53.3\%. However, FT requires a full retraining cycle and may induce representation drift, which can harm previously stable service knowledge.

Compared with GSR, our method consistently improves Recall@5 by 24.8\%, 11.4\%, and 48.0\% on Mashup, Java, and Python, respectively. Compared with FT, our approach achieves gains of 11.0\%, 15.8\%, and 30.0\% across the three datasets.  On Mashup, our method improves Recall@5 from FT’s 0.617 to 0.685, an 11.0\% relative gain, and mAP@10 from 0.700 to 0.760, an 8.6\% improvement. On Java, Recall@5 increases from 0.330 to 0.382, yielding a 15.8\% gain. On Python, Recall@5 improves from 0.287 to 0.373, a 30.0\% gain, while mAP@10 increases from 0.292 to 0.402, corresponding to a substantial 37.7\% improvement. \add{In particular, the improvement in mAP suggests that the proposed method ranks relevant services higher in the recommendation list, while the improvement in Precision indicates that the top-ranked recommendations are more accurate. The gains vary across datasets because the datasets differ in service diversity and the semantic gap between queries and target services.}

\add{Compared with RAG and Agent-based approaches, \tool consistently achieves superior performance, indicating that inference-time adaptation alone is insufficient. Model editing improves knowledge alignment, reducing outdated or incorrect service mappings.
Constrained decoding enforces structural validity, preventing invalid service combinations. These complementary effects jointly lead to improved recommendation accuracy.} These consistent gains indicate that our approach effectively outperforms baseline methods in both effectiveness and robustness.

\begin{tcolorbox}
\textbf{Finding for RQ1.}
\tool delivers state-of-the-art performance on all datasets, achieving a 25.9\% average improvement in Recall@5, which confirms its ability to incorporate new knowledge while maintaining accuracy and robustness comparable to full-data fine-tuning.
\end{tcolorbox}

\subsection{RQ2. Ablation Study and Hyperparameter Analysis}

\begin{figure}[!htbp]
    \centering
    \includegraphics[width=0.44\textwidth]{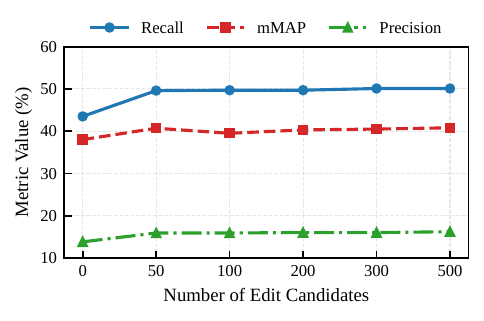}
    \caption{Effect of data scale on Recall@5.}
    \label{fig:recall5_datascale}
\end{figure}

\begin{figure}[!htbp]
    \centering
    \includegraphics[width=0.44\textwidth]{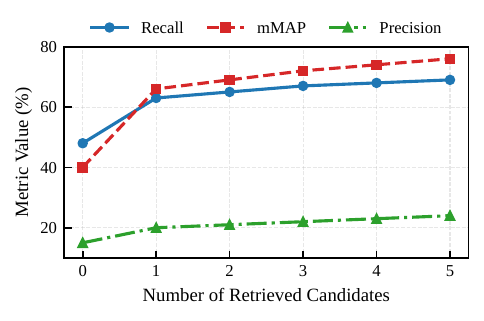}
    \caption{Effect of the number of retrieved candidates on Recall@5. 
   }
    \label{fig:retrieval_number_recall5}
\end{figure}

We conduct ablation experiments based on Qwen2.5-7B to analyze the contribution of each module in our framework, with all results reported at @5. 

\paragraph{Effect of Model Components}
Table~\ref{tab:ablation_5} summarizes the impact of removing individual modules.
Removing the retrieval module leads to a notable drop in Recall and mAP, indicating that high-quality candidate retrieval is essential for narrowing the search space.
Without constrained decoding, performance degrades drastically across all metrics, demonstrating its critical role in enforcing valid and stable predictions.
The editing module further improves performance by refining retrieved candidates, yielding consistent gains in Recall and Precision.
The full model achieves the best results across all metrics, confirming the complementary nature of these components.

\begin{table}[t]
\centering
\caption{Ablation study on the effects of different modules based on the Qwen2.5-7B (metrics @5).}
\begin{tabular}{lccc}
\toprule
\textbf{Model Variant} & 
\textbf{Recall} & \textbf{Precision} & \textbf{mAP} \\
\midrule
w/o Retrieval Module & 52.1 & 17.4 & 56.6 \\
w/o Constrained Decoding & 22.4 & 8.9 & 27.2 \\
w/o Editing Module &  66.3 & 22.2 & 74.0\\
\textbf{Full Model (All)} & \textbf{68.5} & \textbf{23.3} & \textbf{76.0} \\
\bottomrule
\end{tabular}
\label{tab:ablation_5}
\end{table}

\paragraph{Effect of Editing Scale}
Fig.~\ref{fig:recall5_datascale} shows the effect of varying the number of edit candidates.
Recall and mAP increase as the editing scale grows, but the gains become marginal once the scale exceeds 100, indicating clear saturation.
Precision remains relatively stable throughout.
This suggests that the editing module mainly improves coverage by exploring a richer candidate space, with diminishing returns beyond a moderate scale.

\paragraph{Effect of Retrieved Candidates}
Fig.~\ref{fig:retrieval_number_recall5} illustrates the influence of the number of retrieved candidates.
Performance consistently improves as more candidates are retrieved, especially in Recall and mAP, before gradually saturating.
This indicates that the retrieval module effectively captures relevant candidates at small scales, while constrained decoding and editing mitigate noise introduced by additional candidates.

\begin{figure}[!htbp]
    \centering
    \includegraphics[width=0.9\linewidth]{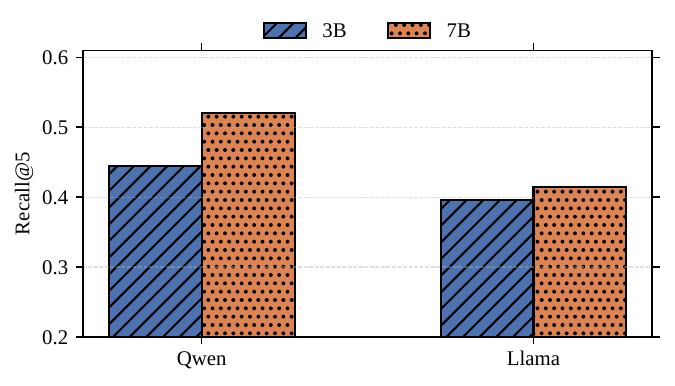}
    \caption{Effectiveness of Different Base Models. }
    \label{fig:recall_base_models}
\end{figure}

\paragraph{Effect of Different Base Models}
We evaluate the impact of different base LLMs on service recommendation performance, as shown in Fig.\ref{fig:recall_base_models}.
Increasing model size from 3B to larger variants, 7B for Qwen and 8B for LLaMA-3, consistently improves Recall@5, demonstrating the benefit of LLMs in representing complex service knowledge.
Compared with LLaMA, Qwen achieves higher recall at both scales, indicating better alignment with the service recommendation task.
Notably, Qwen with 7B achieves the highest performance among all evaluated models, highlighting its advantage in balancing model capacity and knowledge adaptability.
Therefore, we select Qwen with 7B as the base model for the remainder of our experiments.

\begin{tcolorbox}
\textbf{Finding for RQ2.}
Our ablation results demonstrate that each module is critical to the overall effectiveness of the framework.
Retrieval and constrained decoding are indispensable for stable and accurate inference.
\end{tcolorbox}

\begin{figure*}[!htbp]
    \centering
    \begin{minipage}[t]{0.32\linewidth}
        \centering
        \includegraphics[width=\linewidth]{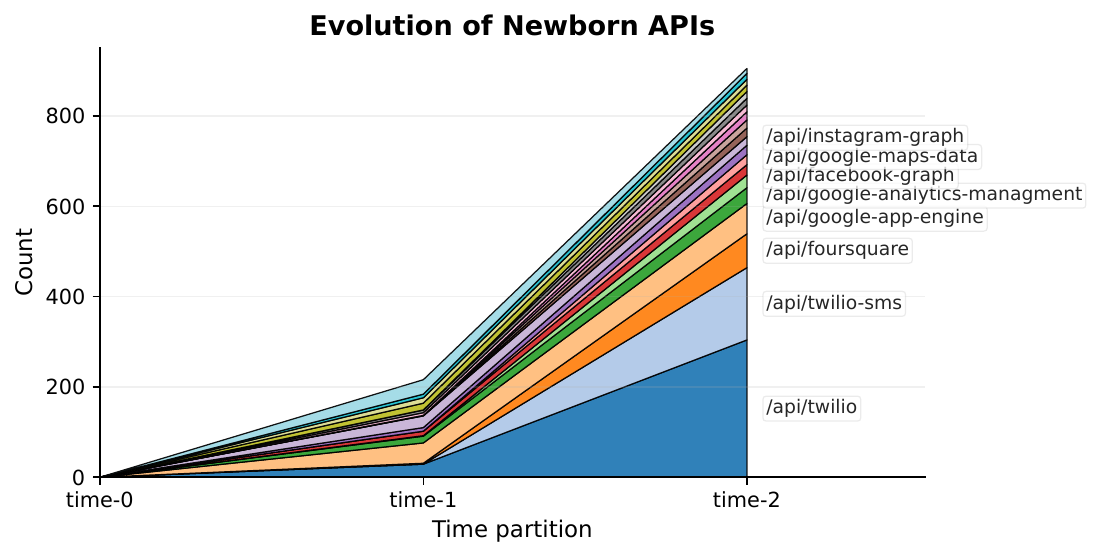}
        \caption*{(a) Newborn}
    \end{minipage}\hfill
    \begin{minipage}[t]{0.32\linewidth}
        \centering
        \includegraphics[width=\linewidth]{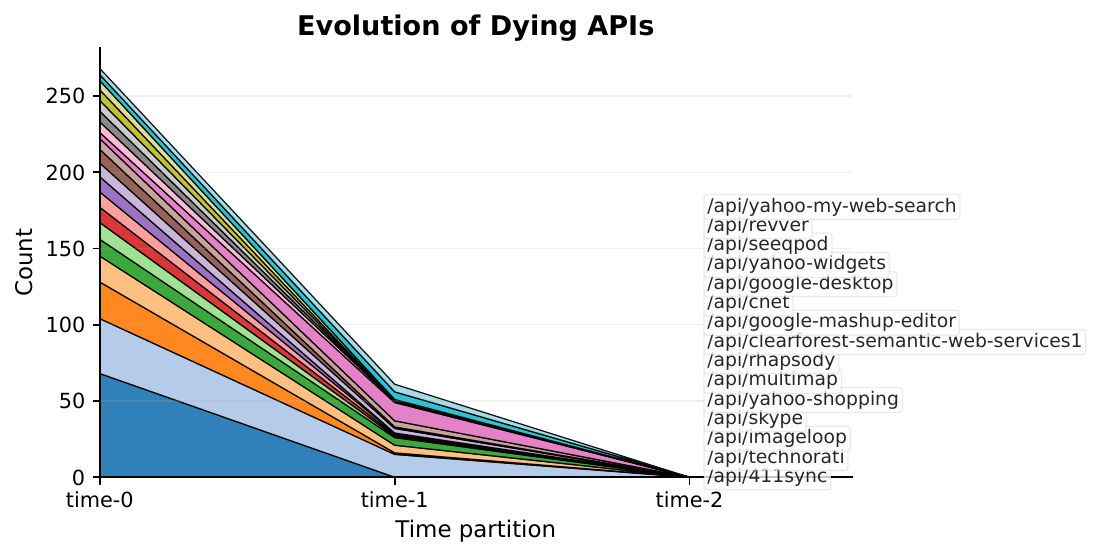}
        \caption*{(b) Dying}
    \end{minipage}\hfill
    \begin{minipage}[t]{0.32\linewidth}
        \centering
        \includegraphics[width=\linewidth]{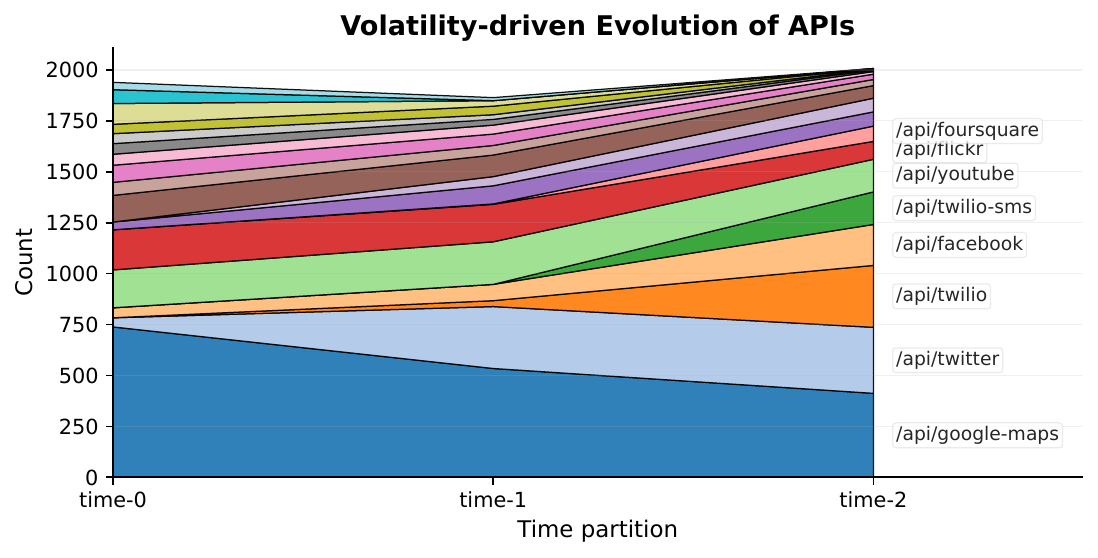}
        \caption*{(c) Volatile}
    \end{minipage}
    \caption{Evolution patterns: newborn, dying, and volatile APIs.}
    \label{fig:api-evolution}
\end{figure*}
\begin{figure}[!htbp]
    \centering
    \includegraphics[width=0.9\linewidth]{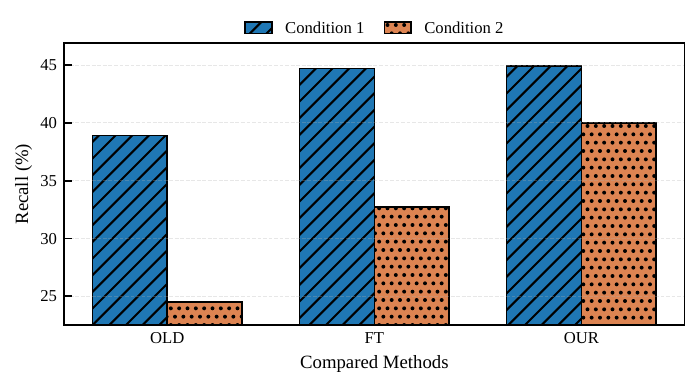}
    \caption{Effectiveness of Knowledge Updating under Service Evolution}
    \label{fig:recall_two_cond}
\end{figure}

\subsection{RQ3. Service Evolution }

\paragraph{Analyzing}
We first analyze the evolution of APIs on the mashup dataset. Complex evolutionary operations, such as API splitting, can be decomposed into two fundamental processes: the disappearance of existing APIs, referred to as dying APIs, and the emergence of new ones, referred to as newborn APIs. Additionally, we examine the evolution of usage distributions, which track the changes in adoption and relative importance of APIs over time, denoted as volatile APIs. The tree evolution patterns are shown in Fig.~\ref{fig:api-evolution}.

\paragraph{Data Design}
We focus on newly introduced services that are invoked more than twice\add{, ensuring that one instance is available for training and at least one for testing.} For each such service, one instance is randomly selected for the training set, while the remaining instances are used for testing. All other services are included exclusively in another test set for testing knowledge preservation. 
This setup evaluates whether knowledge editing is effective on the targeted services and whether it preserves previously learned capabilities on untargeted services.
For dying APIs, we apply constrained decoding during inference.
To emphasize temporal generalization, the base model is trained on data from time $t_0$, and knowledge editing is performed using data from time $t_1$, yielding an evolved model.

\paragraph{Experimental Design}
To validate the effectiveness of our approach, the dataset is organized into a simple format. Each data sample consists of two parts: (1) a model input, which is a textual Description, and (2) a model output, which specifies the corresponding Output Services. This unified input–output structure is used consistently for both training and evaluation.
We compare the performance of different methods under two evaluation conditions in terms of  Recall@5.
OLD denotes the model trained on time0 data and directly evaluated on time2 data without any adaptation.
FT represents a fine-tuning baseline, where one sample is extracted for each new service from the time2 data and used for fine-tuning the model.
\tool denotes our proposed method, where the retrieval module is removed.
For evaluation, Condition 1 corresponds to testing on the full time2 test set, while Condition 2 is a more challenging setting in which every test instance in time2 must contain at least one new service. \add{The improvements under this setting can be interpreted as evidence that model editing enhances the model's coverage of newly introduced services and improves robustness to evolving service distributions.}

\paragraph{Results}
Several observations can be drawn from the results, \add{as shown in Fig.~\ref{fig:recall_two_cond}}. First, the OLD method exhibits a substantial performance drop under Condition 2, indicating that a model trained solely on historical data (time 0) struggles to generalize to newly introduced changes. Second, the FT approach improves performance in both conditions, demonstrating that incorporating a small amount of new-service data via fine-tuning can partially alleviate the distribution shift. However, its overall gains remain limited.
In contrast, OUR method consistently achieves the best performance under both conditions, with particularly notable improvements in Condition 2. This suggests that our approach is more effective at handling the challenges posed by new services and exhibits stronger robustness and generalization ability, even without relying on a retrieval module.

\begin{tcolorbox}
\textbf{Finding for RQ3.}
The results demonstrate that the proposed method significantly outperforms direct fine-tuning and provides a more reliable solution for adapting to evolving service distributions over time.
\end{tcolorbox}

\subsection{RQ4. Experiments in Code Scenarios}

\begin{figure}
    \centering
    \includegraphics[width=0.99\linewidth]{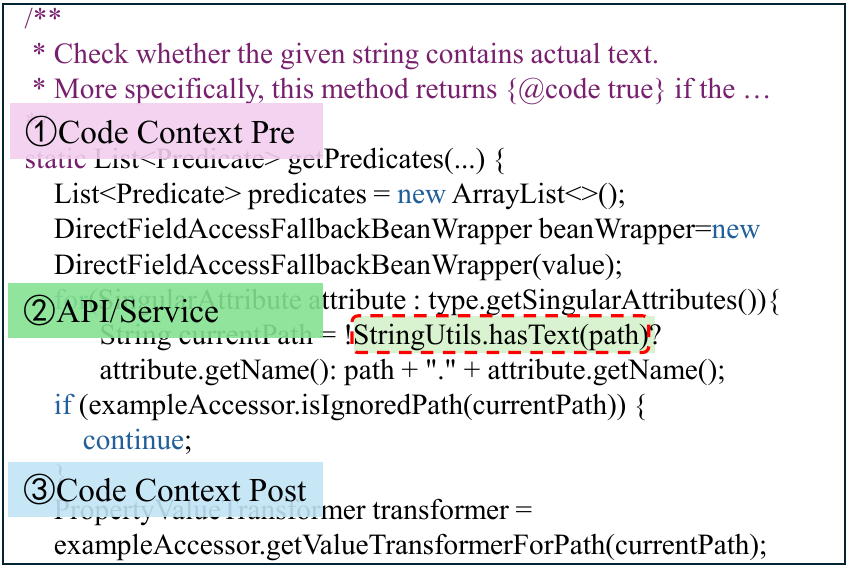}
    \caption{Case Example for Service Recommendations in Code Scenarios. The figure illustrates (\ding{172}) pre-extracted code context,
(\ding{173}) API/service attributes, and (\ding{174}) the resulting code context after service insertion.}
    \label{fig:case}
\end{figure}

\paragraph{Data Design}
We adopt the dataset from Wang et al.\cite{chaozheng2024}, which includes 4,146 entries of API usage in 3,136 code files, covering 176 diverse APIs sourced from 853 popular Java projects. We construct a low-resource training set by selecting one usage instance per API, which contains 176 samples \add{as editing knowledge} in total, while using the remaining instances for evaluation. Function names are extracted from each project and serve as recommendation candidates. Given a code context in which function names are removed, the task is to recommend relevant function names based on the remaining code context, as shown in Fig.\ref{fig:case}.

\begin{table}[!htbp]
\centering
\caption{Performance comparison of different approaches in Code Scenario based on the Qwen2.5-7B (metrics @5).}
\label{tab:code_performance}
\begin{tabular}{lccc}
\toprule
Method & Recall & Precision & mAP \\
\midrule
\multicolumn{4}{l}{\textit{Pre+Post Code Context}} \\
\midrule
Original & 1.8 & 1.7 & 1.5 \\
FT       & 57.1 & \textbf{56.1} & 46.1 \\
\tool    & \textbf{70.6} & 16.9 & \textbf{58.0} \\
\midrule
\multicolumn{4}{l}{\textit{Pre Code context}} \\
\midrule
Original & 5.8 & 6.2 & 5.2 \\
FT       & 54.4 & \textbf{53.9} & 42.1 \\
\tool    & \textbf{70.6} & 14.9 & \textbf{55.1} \\
\bottomrule
\end{tabular}
\end{table}

\paragraph{Experimental Design}
We evaluate the performance comparison of different approaches 
under the code-only scenario, evaluated using Recall, Precision, and mAP. 
We consider two code context settings: (i) \emph{Pre+Post Code Context}, where both the pre- and post-level code surrounding the target API call is available, and (ii) \emph{Pre Code Context}, where only the upper-level code context is provided.

\paragraph{Results}
As shown in Table~\ref{tab:code_performance}, across both settings, the proposed method consistently outperforms the Original 
and FT baselines. In particular, under the Upper+Lower Code Context, 
\tool achieves a Recall@5 of 70.6\% and an mAP@5 of 58.0\%, substantially 
ssurpassing FT, which attains 57.1\% Recall and 46.1\% mAP. Similar trends are 
observed in the more challenging Upper Code Context setting, indicating that our 
approach remains robust even when the available code context is limited.

\begin{tcolorbox}
\textbf{Finding for RQ4.}
Our approach achieves the best performance in the code setting, and more information does not necessarily lead to better results.
\end{tcolorbox}

\subsection{RQ5. Analysis of Constrained Decoding}

\add{To understand the interaction between FA-based constrained decoding and the model’s probability distribution, we conduct a quantitative analysis of probability cost, entropy dynamics, and validity–probability tradeoff, using the mashup dataset.}

\paragraph{Probability cost.}
We measure the impact of constrained decoding on the model’s confidence by computing the average probability cost. 
For each decoding step $t$, let $p_{\text{raw}}(w_t)$ denote the probability of the token selected by unconstrained decoding, and $p_{\text{FA}}(w_t)$ denote the probability of the token selected under FA constraints. 
\begin{figure}
    \centering
    \includegraphics[width=0.86\linewidth]{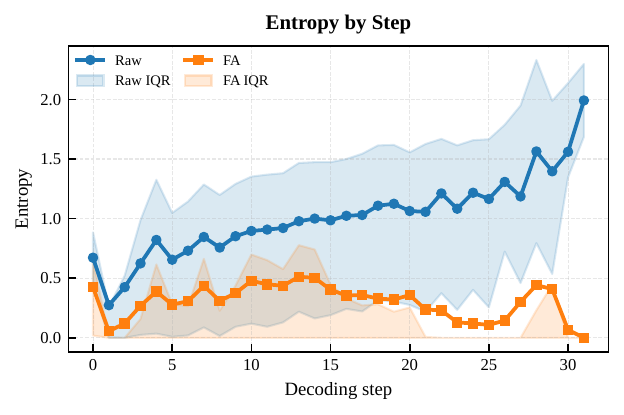}
    \caption{Entropy comparison across decoding steps. A narrower IQR indicates more stable and consistent generation behavior.}
    \label{fig:entropy_step}
\end{figure}
The probability cost is defined as the average log-probability difference, as shown in Eq.~\ref{eq:log_prob}.
\begin{equation}\label{eq:log_prob}
\Delta = \frac{1}{T} \sum_{t=1}^{T} \left( \log p_{\text{FA}}(w_t) - \log p_{\text{raw}}(w_t) \right),
\end{equation}
where $T$ is the sequence length. A lower $\Delta$ value indicates a larger deviation from the model’s original preference.
As shown in Fig.~\ref{fig:entropy_step}, the shaded area represents the interquartile range (IQR), which captures the variability of entropy across samples. FA-based constrained decoding consistently reduces entropy compared to unconstrained decoding. 
This suggests that constraints effectively narrow the candidate token space, leading to more deterministic generation.

\begin{figure}
    \centering
    \includegraphics[width=0.86\linewidth]{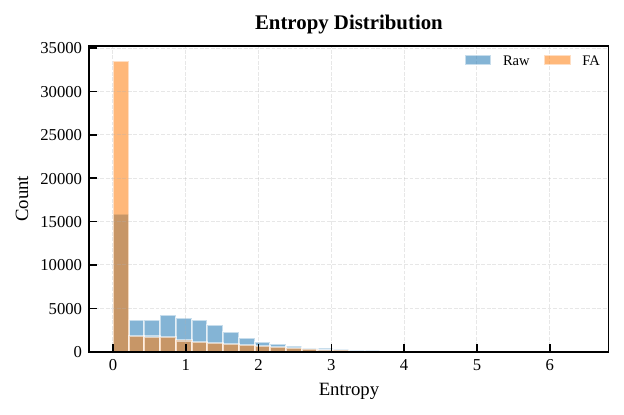}
    \caption{Entropy distribution under constrained decoding. The distribution toward lower entropy values indicates a global reduction in uncertainty.}
    \label{fig:entropy_dist}
\end{figure}

\paragraph{Entropy analysis.}
We further analyze the entropy of the token distribution at each decoding step, defined as $H_t = -\sum_{w} p(w)\log p(w)$, to measure uncertainty during generation. 
As shown in Fig.~\ref{fig:entropy_dist}, Constrained decoding shifts the entropy distribution toward lower values, indicating a global reduction in uncertainty during generation. 
This confirms that FA constraints systematically restrict the model’s output space.

\begin{figure}
    \centering
    \includegraphics[width=0.86\linewidth]{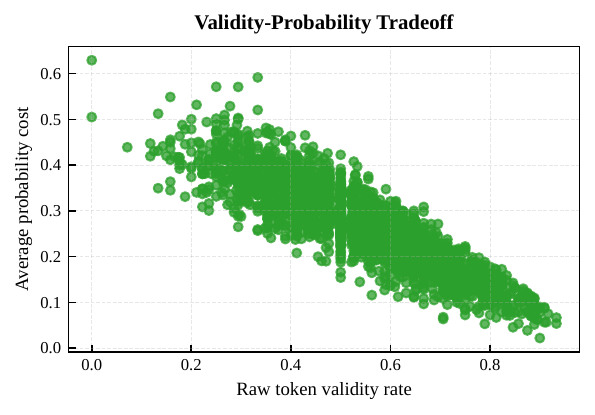}
    \caption{Validity–probability tradeoff.  FA-based constrained decoding consistently reduces entropy by restricting the candidate token space. }
    \label{fig:tradeoff}
\end{figure}
\paragraph{Validity–probability tradeoff.}
We analyze the tradeoff between output validity and probability cost. The validity rate measures the proportion of generated sequences that satisfy structural constraints derived from the Trie-based service lexicon.
As shown in Fig.~\ref{fig:tradeoff}, we observe a clear negative correlation, where as the raw token validity rate decreases, the average probability cost increases. For example, the validity rate is 0.2, and the corresponding cost is around 0.5. 
This indicates that constrained decoding may suppress high-probability tokens when they violate structural constraints, leading to a local reduction in log-probability.

\begin{tcolorbox}
\textbf{Finding for RQ5.}
FA-based constraints reshape the decoding distribution by pruning invalid high-probability regions, effectively projecting the model’s original distribution onto a structurally valid subspace, thereby aligning generation with task constraints.
\end{tcolorbox}
\section{Related Work}
\label{sec:related}

Service recommendation addresses the challenge of selecting, matching, and composing appropriate services to meet dynamic user needs. At the same time, model editing provides efficient mechanisms for updating a model’s internal knowledge without full retraining. 
Building on these advances, service evolution extends the adaptation of services to changing environments and requirements.

\subsection{Service Recommendation}

Research on service recommendation has evolved from rule-based and semantic matching methods to graph-based and deep learning approaches. 
Early rule-based methods relied on manually defined heuristics, business rules, or attribute constraints to identify suitable services, which were simple but lacked scalability~\cite{li2019pre}. 
Semantic matching methods introduced ontology reasoning and semantic similarity metrics to measure the relatedness between user requirements and service descriptions, improving interpretability but dependent on manually crafted knowledge bases~\cite{chen2009building}.

With the development of deep learning, representation learning methods were applied to map textual and behavioral service data into latent vector spaces using models such as Word2Vec or BERT~\cite{fan2023service, wu2021mashup}. 
More recently, Graph Neural Network (GNN) models are introduced to jointly capture both the structural and semantic characteristics of service graphs through message passing, achieving improved accuracy, robustness, and generalization in service recommendation tasks ~\cite{liu2023dysr}.
As service ecosystems grow more complex, the above-mentioned approaches~\cite{liu2023dysr, fan2023service, wu2021mashup, li2019pre, chen2009building} perform recommendations through graph traversal, similarity propagation, or ranking algorithms, enabling structural awareness but often relying on predefined relationships.

\add{Another line of work focuses on generating executable code that integrates API calls. Recent surveys~\cite{wang2024systematic} and models such as CodeLlama and DeepSeek-Coder show strong performance in code synthesis, including API usage and dependency handling. These methods learn invocation patterns from large-scale code corpora and can generate syntactically correct and functionally plausible code snippets. 
Beyond code generation, several studies further explore API recommendation and orchestration, where models first select relevant APIs and then compose them into executable workflows~\cite{fan2026llms}. This paradigm bridges natural language understanding with program synthesis and enables end-to-end task automation. 
}

\begin{table*}[t]
\centering
\small
\caption{Comparison of different approaches for handling evolving service environments.}
\label{tab:comparison}
\begin{tabular}{lccccc}
\toprule
\textbf{Approach} & \textbf{Knowledge Update} & \textbf{Composition Reliability} & \textbf{Constraint Awareness} & \textbf{Context Overhead} & \textbf{Cost} \\
\midrule
Fine-tuning & Global & High & Low & Low & High \\
RAG & External & Low & Low & High & Medium \\
Agent-based & External & Medium & Low & High & Medium \\
Model Editing & Local & High & Low & Low & Low \\
\textbf{Ours} & Local and structured & High & High & Low & Low \\
\bottomrule
\end{tabular}
\end{table*}

\add{\subsection{LLM-based Tool Use and Agent Systems}
We identify three closely related research directions, including LLM-based API invocation or tool use, and agent-based orchestration systems. }

\add{
Recent studies have explored leveraging LLMs for API selection and tool invocation. Toolformer~\cite{schick2023toolformer} enables models to learn when and how to call external tools through self-supervised signals, while ReAct~\cite{yao2022react} integrates reasoning and acting to construct tool-use trajectories iteratively. Gorilla~\cite{patil2024gorilla} further focuses on API selection and invocation by aligning model outputs with API documentation, improving the accuracy of API calls.  More recent systems, including API-Bank~\cite{li2023api}, ToolBench~\cite{guo2024stabletoolbench}, and SWE-bench~\cite{jimenez2023swe}, provide large-scale benchmarks and frameworks for evaluating service use capabilities in realistic scenarios. In software engineering, identifying significant limitations in evaluating LLM-based systems, particularly in realistic tool-use, execution correctness, and complex decision-making scenarios are also systematically reviewed~\cite{koohestani2025benchmarking}.
}

\add{Agent frameworks, such as ReAct~\cite{yao2022react}, HuggingGPT~\cite{shen2023hugginggpt}, Agentic Context Engineering~\cite{zhang2026agentic}, and Harness Engineering~\cite {qi2026harnessing}, further extend this paradigm by dynamically selecting and invoking tools in multi-step reasoning processes. ReAct~\cite{yao2022react} integrates reasoning and acting to iteratively construct tool-use trajectories, while Toolformer~\cite{schick2023toolformer} enables models to learn when to invoke external tools through self-supervised signals. More recent LLM-based autonomous agents~\cite{wang2026openclaw} further generalize this paradigm by enabling iterative planning, memory augmentation, and tool orchestration in complex environments. Meanwhile, emerging explainability research emphasizes the role of concept polysemanticity and proposes entropy-based metrics to characterize uncertainty in model interpretations~\cite{yu2025coe}.}

\add{Existing approaches rely on external retrieval or tool use at inference time. In contrast, \tool performs persistent and localized knowledge updates, yielding more stable and valid service recommendations. 
These paradigms are complementary, and \tool can be integrated with agent-based frameworks. The comparison is shown in Table~\ref{tab:comparison}. 
}

\subsection{Model Editing }
Model editing aims to modify a trained model’s internal knowledge in a localized and efficient manner without retraining from scratch.
It has become an important topic for maintaining and updating LLMs as world knowledge evolves.

\paragraph{Fine-tuning-based Approaches}
Early attempts perform additional fine-tuning to inject new facts or correct wrong ones~\cite{ovadia2024fine}. Typical methods include task-specific retraining or parameter-efficient fine-tuning, e.g., LoRA~\cite{hu2022lora}. While simple and intuitive, these methods are costly and prone to catastrophic forgetting, i.e., interference with unrelated knowledge.

\paragraph{Meta-learning-based Approaches}
Meta-learning methods train an auxiliary ``editor'' network to predict how to adjust model parameters for new edits.
Representative works include MEND~\cite{mitchellfast} and IKE~\cite{zheng2023can}. Such approaches allow fast, one-shot editing at inference time. However, these methods often require accessing a large portion of the model’s internal activations and parameters during editing, resulting in substantial computational overhead and high sample complexity.

\paragraph{Locate-then-Edit Approaches}
A prominent line of work locates the parameters responsible for a fact and performs targeted modification. ROME uses causal tracing to identify and apply rank-one updates to MLP weights, while MEMIT extends this idea to simultaneous multi-fact editing~\cite{meng2022mass}. These methods offer high locality and interpretability, yet may lose stability under sequential or large-scale edits.

\paragraph{Extension-based Approaches}
Another strategy involves adding external modules or memory components to encode edits without modifying the base model, e.g., wise~\cite{wang2024wise}. Examples include dynamic adapter layers or external key-value memories. Such methods are safe and reversible but introduce additional parameters and possible inconsistencies between the edited and base representations.

\begin{table}[t]
\centering
\caption{Memory characteristics of different adaptation strategies.}
\label{tab:memory}
\begin{tabular}{lccc}
\toprule
Method & Grad. Train & Update Scope & Memory \\
\midrule
Full Fine-tuning & Yes & Full model & High \\
LoRA / Adapter  & Yes & Partial modules & Low \\
ROME (Editing)  & No  & Localized weights & Very Low \\
\bottomrule
\end{tabular}
\end{table}

\subsection{Software Service Evolution}

Software service evolution addresses the adaptation of services over time in response to changing requirements, technologies, and environments~\cite{fan2025user}. 
Early work focused on versioning and lifecycle management of web services, proposing mechanisms for schema evolution, backward compatibility, and service deprecation~\cite{liu2021data}. 
Subsequently, research shifted towards service composition and dynamic adaptation, where services are re-bound at runtime according to evolving user needs, context, or QoS requirements.

More recently, service evolution is studied in micro-service and cloud-native architectures, emphasising continuous deployment, automated refactoring, and dependency management across heterogeneous platforms~\cite{cerny2025towards}. 
Techniques such as service migration, container orchestration, and service mesh are applied to support resilient evolution and rollback strategies~\cite{farkiani2024service}. 

Despite these advances, most existing studies focus on structural or operational aspects of service evolution, while the knowledge and behaviour of intelligent service models themselves are rarely considered as first-class evolvable entities. 
Therefore, in this paper, we explore a model-level evolution paradigm for service recommendation, viewed from a knowledge-update perspective, where service behaviour can be dynamically updated or corrected through model editing techniques.





\section{Discussion}

This work presents a novel attempt to bridge model editing and software service evolution, offering a novel perspective on adaptive service recommendation.

\subsection{Advantages}


First, the proposed framework formulates service evolution as a localized knowledge update problem, enabling targeted modification of model behavior without affecting unrelated knowledge. 
This locality property allows the system to adapt to evolving service ecosystems while preserving previously acquired associations, which is essential for maintaining behavioral consistency under continuous evolution.
Second, the FA-based constrained decoding mechanism provides explicit structural and semantic guarantees during generation. 
By enforcing service availability and composition constraints at decoding time, the framework ensures that generated service sequences remain valid and well-formed, even when the underlying service set changes dynamically.
\add{The additional cost mainly comes from constructing the valid token set \(V_{\text{valid}}\) and masking logits at each step, with complexity \(O(|V_{\text{valid}}|)\), where \(V\) denotes the full vocabulary. Since \(|V_{\text{valid}}| \ll |V|\), the overhead remains limited.}
More importantly, the tight integration of knowledge editing and constraint-guided reasoning yields a stable adaptation mechanism that separates knowledge acquisition from validity enforcement. 
This separation enables incremental updates to evolving service knowledge while systematically limiting regression on unaffected services, offering a principled balance between adaptability and stability.
Overall, these advantages make the proposed approach particularly suitable for long-running LLM-based service-dependent systems that must operate reliably in the face of continuous service evolution.

\subsection{Limitations}
Although the proposed FA-augmented decoding effectively enforces structural constraints, it may occasionally conflict with the model’s native probability distribution.
In practice, the model occasionally produces tokens, such as separators, brackets, or fused segments connecting two adjacent services, with high confidence, that contradict the FA constraints. These forced rejections result in abrupt declines in local log-probability and can compromise the smoothness of generation. 
As a consequence, constrained decoding can impair the model’s ability to terminate generation automatically, 
resulting in reduced precision. 
Future work may explore adaptive logit reweighting or constraint relaxation to reduce these decoding–model inconsistencies.
Furthermore, despite the strong potential of model editing, \add{its effectiveness depends on the availability and scale of editing data}. As a result, the system may still encounter challenges such as incomplete generalization \add{, particularly under large-scale or frequent updates, as well as potential} catastrophic forgetting when integrating new knowledge. \add{Our approach is particularly suitable for scenarios where service evolution is frequent but localized, allowing efficient knowledge updates via model editing without full retraining.}

\subsection{Threats to Validity}

\add{Our study may be subject to several threats to validity.}

\add{\textbf{Internal validity.} The effectiveness of our approach may be influenced by implementation choices, including the selection of the editable layer, hyperparameter settings such as learning rate and KL coefficient, and the design of constrained decoding. 
To mitigate this threat, we adopt widely used configurations reported in prior work, perform sensitivity analysis on key hyperparameters, and keep the editing strategy consistent across all experiments.}

\add{\textbf{External validity.}
Our evaluation is conducted on datasets derived from ProgrammableWeb and code/query-based API data. Although these datasets cover diverse services, they may not fully represent all real-world service ecosystems, especially large-scale or rapidly evolving environments.  To mitigate this threat, we evaluate our approach across these multiple datasets with different characteristics and include both static and evolving settings, and design chronological splits to better approximate evolution.}

\add{\textbf{Construct validity.} We adopt standard ranking metrics, including Recall, Precision, and mAP, to evaluate recommendation performance. These metrics primarily assess accuracy and ranking quality, but may not fully reflect other important aspects, such as execution correctness, robustness, or the long-term stability of generated service compositions. 
To mitigate this threat, we incorporate constrained decoding to enforce structural validity during generation, and further analyze validity rate, entropy, and probability cost (RQ5) to provide additional evidence on generation quality and model behavior. 
}

\subsection{Implications for Software Engineering Practice}

Our work lies at the intersection of software engineering and service computing, enabling the services to meet users’ requirements.
Our findings offer several practical implications for the design and maintenance of intelligent service recommendation systems in evolving software ecosystems.

First, our results suggest that model editing can serve as a lightweight and cost-effective alternative to full retraining when adapting LLM-based systems to frequent service changes. In practice, service repositories evolve continuously, with APIs being introduced, deprecated, or modified at a pace that makes repeated retraining impractical. By enabling localized knowledge updates, \tool allows practitioners to incorporate service changes incrementally while preserving previously learned behaviors.
Second, our study highlights the importance of enforcing structural and semantic constraints during generation. While LLMs exhibit strong generative capabilities, unconstrained decoding may produce invalid, non-existent, or redundant service recommendations, which can undermine system reliability in real-world deployments. 
Third, the experimental results indicate that combining model editing, constrained decoding, and information retrieval provides a complementary balance between adaptability and stability. Model editing improves coverage of newly introduced services, while constrained decoding mitigates the instability and error propagation caused by frequent updates. 
Meanwhile, the retrieval module helps align the model with evolving service data distributions by dynamically exposing relevant and up-to-date service candidates at inference time.

\section{Conclusion}

In this paper, we studied the problem of service recommendation in evolving service ecosystems and highlighted the limitations of existing approaches that rely on static training data and costly retraining procedures.
To address service evolution from a knowledge-update perspective, we proposed \textbf{\tool}, a generative LLM-based constrained recommendation framework that integrates model editing and finite-automata(FA)-based constrained decoding.
By adopting a locate--then--edit paradigm, \tool enables efficient and localized updates of evolving service knowledge without requiring full model retraining, allowing the recommendation model to remain aligned with newly introduced, updated, or deprecated services.
To further ensure the validity and executability of generated recommendations, we introduced an FA-based constrained decoding mechanism with deduplication, which effectively enforces structural and semantic constraints and mitigates hallucinated or redundant service outputs.
Extensive experiments on real-world service datasets demonstrate that our approach outperforms baseline methods.

\section*{Acknowledgment}
This work is supported by the National Natural Science Key Foundation of China, grant No.62032016, the National Natural Science Foundation of China, grant No. 62472126, the Shandong Provincial Natural Science Foundation for Young Scholars under Grant No. ZR2025QC2234Z,  and by the Shandong Agriculture and Engineering University Start-Up Fund for Talented Scholars under grant No. 2025GCCZR-29.

\balance
\bibliographystyle{IEEEtran}
\bibliography{reference}

\end{document}